\documentstyle[12pt]{article}

\newcommand{\newc}{\newcommand} 
\newc{\ra}{\rightarrow} 
\newc{\lra}{\leftrightarrow} 
\newc{\beq}{\begin{equation}} 
\newc{\eeq}{\end{equation}} 
\newc{\barr}{\begin{eqnarray}} 
\newc{\earr}{\end{eqnarray}} 

\begin{document} 
\begin{center}
{\bf Modulation effect in the differential rate for Supersymmetric  Dark Matter 
detection. }

\vspace{5mm} 

J. D. VERGADOS 

\vspace{3mm} 

Theoretical Physics Section, University of Ioannina, GR-45110, Greece. 
\end{center}

\vspace{3mm} 

\abstract{\it The modulation effect in the direct detection of supersymmetric 
Cold Dark Matter
(CDM) particles is investigated.
 It is shown that, while normally the modulation effect in the total
event rate is small, $\leq 5\% $, in some special cases it becomes much 
larger. It also becomes more pronounced in the
differential event rate. It may thus be exploited to discriminate against
background.} 

\bigskip
\bigskip
\centerline{\bf 1. INTRODUCTION }
\bigskip
In this work we will study the differential modulation effect
of the event rate for detecting supersymmetric dark matter, i.e. its
variation with respect to the energy transferred to the nucleus,
due to the Earth's motion.

There is now ample evidence that most of the matter of the Universe is
non luminous, i.e. dark \cite{Jungm} and is composed of two components.
One is the Hot Dark Matter (HDM) component consisting of particles which 
were relativistic at freeze out, while the other is the Cold Dark Matter (CDM)
composed of particles which were non relativistic. 
There are many arguments supporting the fact that the CDM is at least  
$60\%$~\cite{COBE}. There are two interesting cold dark
matter candidates:
i) Massive Compact Halo Objects (MACHO's) and
ii) Exotic Weakly Interacting Massive Particles (WIMP's).
Since the MACHO's cannot exceed $40\%$ of the CDM component~\cite{Jungm,Benne},
there is room for an exotic candidate. The most natural one is associated
with supersymmetry, i.e. the lightest supersymmetric particle (LSP).

The most interesting possibility to directly detect the LSP \cite{Jungm,
JDV}
is via the recoiling
of a nucleus (A,Z) in the process:
\begin{equation}
\chi \, +\, (A,Z) \, \to \, \chi \,  + \, (A,Z)^* 
\end{equation}
($\chi$ denotes the LSP). In the above process only the elastic channel is open
since the energy of the LSP is too low to excite the nucleus. In computing the 
event rate for the above process one proceeds with the following steps:

1) Write down the effective Lagrangian at the elementary particle 
(quark) level in the framework of supersymmetry as described 
in Refs. \cite{Jungm,JDV} .

2) Go from the quark to the nucleon level using an appropriate quark 
model for the nucleon. Special attention must be paid to the 
scalar couplings, which dominate the coherent part of the cross section
and the isoscalar axial current, which
strongly depend on the assumed quark model \cite{JDV,
Dree} .

3) Compute the relevant nuclear matrix elements ~\cite{KVprd,Ress,
Dimit,Engel,Nikol} 
using as reliable as possible many body nuclear wave functions.

4) Calculate the modulation of the event rate due to the Earth's
revolution around the sun \cite{KVprd,Druk}.

There are many popular targets~\cite{Smith,Primack,Berna}
for LSP detection as e.g. $^{19}F$,
$^{23}Na$, $^{27}Al$, $^{29}Si$, $^{40}Ca$, $^{73,74}Ge$, $^{127}I$, 
$^{207}Pb$, etc.

In a previous paper \cite{KVprd} we computed the modulation effect h, i.e. 
the oscillation amplitude of the total event rate (see below for 
its precise definition),
by convoluting with the LSP velocity distribution the event rate, which, among
other
things, depends upon the relative velocity of the LSP with respect to the Earth.
Assuming a Maxwell Boltzmann distribution \cite{Jungm} of velocities for LSP,
we found that h cannot exceed the value of $ 5\% $ which corresponds to
small momentum transfer. The actual value of h is quite a bit smaller
especially for heavy nuclei and relatively heavy LSP $(m_{\chi} \geq 50 GeV )$.
We showed that, in some cases, the quantity h may become negative suggesting
cancellations between the bins that correspond to small 
and those which correspond to relatively large energy transfers. It is, thus,
possible that in some energy bins the modulation effect  can be larger than the the value of h quoted above.

The event rate
depends on many parameters \cite{JDV}, since 
there exist many contributions to the above process. The most dominant appears
to be the coherent contribution, which arises out of the scalar coupling
originating from Higgs exchange or squark exchange if there exists mixing 
between the L and R squark varieties. It can also arise from the time component of the vector 
current originating from s-quark and Z-exchange. The latter is favored from the
point of view of the couplings but it is suppressed  kinematically 
by factors of $\beta ^{2} \sim 10^{-6} $ owing to the fact that the LSP is a
Majorana particle. Due to its different dependence on the LSP velocity, it   
yields a higher modulation effect. In addition to the coherent part, especially
for light targets, when the target spin is non zero, one must include the
 axial
current (spin matrix element of the nucleus).\\  
Our main purpose is to calculate the convoluted differential modulation effect  
H, i.e. the ratio of the part of the differential event rate which depends on 
the position of the Earth divided by that which does not (for its 
definition see below). If one considers each of the above mechanisms separately,
H depends only on the LSP mass and the size of the nucleus. Knowledge of H may not be adequate, however, since one needs to know its value in the energy transfer regime
where the event rate is the largest and hopefully measurable. One might also
need the relative differential event rate, i.e. the ratio of 
the differential rate to the total rate. If one considers 
the above three mechanisms separately, the relative differential event rate 
 is independent of the SUSY parameters 
or the structure of the nucleon. It  depends on the nuclear structure
only mildly through the form factors. So  one can make quite accurate
predictions which depend only on the nuclear size, the mass of the LSP and the
low energy cutoff imposed by the detector. 

\bigskip
\centerline{\bf 2. EXPRESSIONS FOR THE RATE}
\bigskip

As we have mentioned in the introduction we only need calculate the
fraction of the differential rate divided by the total rate which
is independent of the parameters of supersymmetry. Thus we are not going
to elaborate here further on these, but refer the reader to the literature \cite
{JDV,KVprd,ref1,ref2}.  For completeness we only give here expressions 
describing the effective Lagrangian obtained in first
order via Higgs exchange, s-quark exchange and Z-exchange. We will use
a formalism which is familiar from the theory of weak interactions, i.e. 
 \beq
{\it L}_{eff} = - \frac {G_F}{\sqrt 2} \{({\bar \chi}_1 \gamma^{\lambda}
\gamma_5 \chi_1) J_{\lambda} + ({\bar \chi}_1 
 \chi_1) J\}
 \label{eq:eg 41}
\eeq
where
\beq
  J_{\lambda} =  {\bar N} \gamma_{\lambda} (f^0_V +f^1_V \tau_3
+ f^0_A\gamma_5 + f^1_A\gamma_5 \tau_3)N
 \label{eq:eg.42}
\eeq
and
\beq
J = {\bar N} (f^0_s +f^1_s \tau_3) N
 \label{eq:eg.45}
\eeq

We have neglected the uninteresting pseudoscalar and tensor
currents. Note that, due to the Majorana nature of the LSP, 
${\bar \chi_1} \gamma^{\lambda} \chi_1 =0$ (identically).
The parameters $f^0_V, f^1_V, f^0_A, f^1_A,f^0_S, f^1_S$  depend
on the SUSY model employed. In SUSY models derived from minimal SUGRA
the allowed parameter space is characterized at the GUT scale by five 
parameters,
two universal mass parameters, one for the scalars, $m_0$, and one for the
gauginos, $m_{1/2}$, as well as the parameters 
$tan\beta $, one of $ A_0 $, or $ m^{pole}_t $  and the
sign of $\mu $ \cite{ref3}. Deviations from universality at the GUT scale
have also been considered and found useful \cite{ref4}. We will not elaborate
further on this point since the above parameters involving universal masses
have already been computed in some models \cite{JDV,KVdubna} and effects
resulting from deviations from universality will be published elsewhere
\cite{WKV} (see also Arnowitt {\it et al} in Ref. \cite{ref4}).

The invariant amplitude in the case of non-relativistic LSP 
can be cast in the form~\cite{JDV}
\barr
|{\cal M}|^2 &=& \frac{E_f E_i -m^2_x +{\bf p}_i\cdot {\bf p}_f } {m^2_x} \,
|J_0|^2 +  |{\bf J}|^2 +  |J|^2 
 \nonumber \\ & \simeq & \beta ^2 |J_0|^2 + |{\bf J}|^2 + |J|^2 
\label{2.1}
 \earr
where $m_x$ is the LSP mass, $|J_0|$ and $|{\bf J }|$ indicate the matrix 
elements of the time and space components of the current $J_\lambda$ 
of Eq. (\ref{eq:eg.42}), respectively, and $J$ represents the matrix 
element of the 
scalar current J of Eq. (\ref{eq:eg.45}). Notice that $|J_0|^2$ is multiplied
by $\beta^2$ (the suppression due to the Majorana nature of LSP mentioned
above). It is straightforward to show that 
\beq
 |J_0|^2 = A^2 |F({\bf q}^2)|^2 \,\left(f^0_V -f^1_V \frac{A-2 Z}{A}
 \right)^2
\label{2.2}
\eeq
\beq
 J^2 = A^2 |F({\bf q}^2)|^2 \,\left(f^0_S -f^1_S \frac{A-2 Z}{A}
 \right)^2
\label{2.3}
\eeq
\beq
|{\bf J}|^2 = \frac{1}{2J_i+1} |\langle J_i ||\, [ f^0_A {\bf
\Omega}_0({\bf q})\,+\,f^1_A {\bf \Omega}_1({\bf q}) ] \,||J_i\rangle |^2 
\label{2.4}
\eeq
with $F({\bf q}^2)$ the nuclear form factor and
\beq 
{\bf \Omega}_0({\bf q})  = \sum_{j=1}^A {\bf \sigma}(j) e^{-i{\bf q} \cdot
{\bf x}_j }, \qquad
{\bf \Omega}_1({\bf q})  = \sum_{j=1}^A {\bf \sigma} (j) {\bf \tau}_3 (j)
 e^{-i{\bf q} \cdot {\bf x}_j }
\label{2.5}  
\eeq
where ${\bf \sigma} (j)$, ${\bf \tau}_3 (j)$, ${\bf x}_j$ are the spin, third
component of isospin ($\tau_3 |p\rangle = |p\rangle$) and coordinate of
the j-th nucleon and $\bf q$ is the momentum transferred to the nucleus.

The differential cross section in the laboratory frame takes the form~\cite{JDV}
\barr
\frac{d\sigma}{d \Omega} &=& \frac{\sigma_0}{\pi} (\frac{m_x}{m_N})^2
\frac{1}{(1+\eta)^2} \xi  \{\beta^2 |J_0|^2  [1 - \frac{2\eta+1}{(1+\eta)^2}
\xi^2 ] + |{\bf J}|^2 + |J|^2 \} 
\label{2.6}
 \earr
where $m_N$ is the proton mass, $\eta = m_x/m_N A$, $ $
$\xi = {\bf {\hat p}}_i \cdot {\bf {\hat q}} \ge 0$ (forward scattering) and  
\beq
\sigma_0 = \frac{1}{2\pi} (G_F m_N)^2 \simeq 0.77 \times 10^{-38}cm^2 
\label{2.7} 
\eeq
The momentum transfer  $\bf q$ is given by
\beq
|{\bf q}| = q_0 \xi, \qquad q_0 = \beta \frac{2  m_x c }{1 +\eta}
\label{2.8} 
\eeq
Some values of $q_0$ (forward momentum transfer) for some characteristic values
of $m_x$ and representative nuclear systems (light, medium and heavy)
are given in Ref. \cite{KVprd}. It is clear from Eq. (\ref{2.8}) that the momentum
transfer can be sizable for large $m_x$ and heavy nuclei ($\eta$ small).

Integrating the differential cross section, Eq. (\ref {2.6}), with respect to the azimuthal
angle we obtain
\barr
d\sigma (u_0,\xi) &=& \sigma_0 (\frac{m_x}{m_N})^2 \frac{1}{(1+\eta)^2} \,
 \{\{ A^2 \, [[\beta^2 (f^0_V - f^1_V \frac{A-2 Z}{A})^2 
\nonumber \\ & + & 
(f^0_S - f^1_S \frac{A-2 Z}{A})^2 \, ]F^2(u_0 \xi ^2) -
\frac{(\xi \beta )^2}{2} \frac{2\eta +1}{(1+\eta)^2}
(f^0_V - f^1_V \frac{A-2 Z}{A})^2 F^2(u_0\xi ^2) ]
\nonumber \\ & + & 
(f^0_A \Omega_0(0))^2 F_{00}(u_0\xi ^2) + 2f^0_A f^1_A \Omega_0(0) \Omega_1(0)
F_{01}(u_0\xi ^2)  \nonumber \\ 
&+& (f^1_A \Omega_1(0))^2 F_{11}(u_0\xi ^2) \, \}\} 2\xi d\xi
\label{2.9}
 \earr
Where
\beq
F_{\rho \rho^{\prime}}(u_0\xi ^2) =  \sum_{\lambda,\kappa}
\frac{\Omega^{(\lambda,\kappa)}_\rho( u_0\xi^2)}{\Omega_\rho (0)} \,
\frac{\Omega^{(\lambda,\kappa)}_{\rho^{\prime}}( u_0\xi^2)}
{\Omega_{\rho^{\prime}}(0)} 
, \qquad \rho, \rho^{\prime} = 0,1
\label{2.10} 
\eeq
The total cross section $\sigma (u_0,\beta)$, whch has been studied previously
(see e.g. \cite{JDV,KVprd}), can be cast in the form
\barr
\sigma &=& \sigma_0 (\frac{m_x}{m_N})^2 \frac{1}{(1+\eta)^2} \,
 \{ A^2 \, [[\beta^2 (f^0_V - f^1_V \frac{A-2 Z}{A})^2 
\nonumber \\ & + & 
(f^0_S - f^1_S \frac{A-2 Z}{A})^2 \, ]I_0(u_0) -
\frac{\beta^2}{2} \frac{2\eta +1}{(1+\eta)^2}
(f^0_V - f^1_V \frac{A-2 Z}{A})^2 I_1 (u_0) ]
\nonumber \\ & + & 
(f^0_A \Omega_0(0))^2 I_{00}(u_0) + 2f^0_A f^1_A \Omega_0(0) \Omega_1(0)
I_{01}(u_0)  \nonumber \\ 
&+& (f^1_A \Omega_1(0))^2 I_{11}(u_0) \, \} 
\label{2.11}
 \earr
The quantities $I_{\rho}$ entering Eq. (\ref{2.9}) are defined as 
\beq
I_\rho(u_0)  =   (1+\rho)u_0^{-(1+\rho)}
  \int_0^{u_0} x^{1+ \rho} \, |F( x)|^2 \,dx,
\qquad \rho = 0,1
\label{2.12} 
\eeq
where $F(u_0 \xi ^2)$ the nuclear form factor and
\beq
u_0 = q_0^2b^2/2
\label{2.13} 
\eeq

The integrals $I_{\rho\rho^{\prime}}$, with $\rho,\rho^{\prime} =0,1$,
result by following the standard procedure of the multipole
expansion of the $e^{-i {\bf q} \cdot {\bf r}}$ in Eq. (\ref{2.5}). One finds
\beq
I_{\rho \rho^{\prime}}(u_0) = 2 \int_0^1 \xi \, d\xi \sum_{\lambda,\kappa}
\frac{\Omega^{(\lambda,\kappa)}_\rho( u_0\xi^2)}{\Omega_\rho (0)} \,
\frac{\Omega^{(\lambda,\kappa)}_{\rho^{\prime}}( u_0\xi^2)}
{\Omega_{\rho^{\prime}}(0)} 
, \qquad \rho, \rho^{\prime} = 0,1
\label{2.14} 
\eeq

For the evaluation of the differential rate, which is the main subject of the
present work, it will be more convenient to use
the variables $ (\upsilon ,u) $ instead of
the variables $ (\upsilon ,\xi) $. Thus we get
\barr
d\sigma (u,\upsilon) &=& \sigma_0 (\frac{m_x}{m_N})^2 \frac{1}{(1+\eta)^2} \,
 \{ A^2 \, [[(\frac{\upsilon}{c})^2 (f^0_V - f^1_V \frac{A-2 Z}{A})^2 
\nonumber \\ & + & 
(f^0_S - f^1_S \frac{A-2 Z}{A})^2 \, ]F^2(u) -
\frac{1}{(\mu _r b)^2} \frac{2\eta +1}{(1+\eta)^2}
(f^0_V - f^1_V \frac{A-2 Z}{A})^2 u F^2(u) ]
\nonumber \\ & + & 
(f^0_A \Omega_0(0))^2 F_{00}(u) + 2f^0_A f^1_A \Omega_0(0) \Omega_1(0)
F_{01}(u)  \nonumber \\ 
&+& (f^1_A \Omega_1(0))^2 F_{11}(u) \, \} \frac{du}{2 (\mu _r b)^2}
\label{2.15}
\earr
\beq
u = q^2b^2/2, \qquad \mu _r = \frac{m_{\chi }}{1 + \eta}
\label{2.16} 
\eeq
where $\mu _r$ is the reduced mass and 
the quantity u is related to the experimentally measurable energy transfer
 Q via the relations
\beq
Q=Q_{0}u, \qquad Q_{0} = \frac{1}{A m_{N} b^2} 
\label{2.17} 
\eeq
Let us now assume that the LSP is moving with velocity $v_z$ with 
respect to the detecting apparatus. Then, the detection rate 
for a target with mass $m$ is given by

\beq
R =\frac{dN}{dt} =\frac{\rho (0)}{m} \frac{m}{A m_N} | v_z | \sigma (u,\upsilon)
\label{2.18}  
\eeq
where $\rho (0) = 0.3 GeV/cm^3$ is the LSP density in our vicinity. 
This density has to be consistent with the LSP velocity distribution 
(see next section).

The differential rate can be written as
\beq
dR = \frac{\rho (0)}{m} \frac{m}{A m_N} | v_z | d\sigma (u,\upsilon)
\label{2.19}  
\eeq
where $d\sigma(u,\upsilon )$ is given by Eq. ( \ref{2.15})

\bigskip
\centerline{\bf 3. CONVOLUTION OF THE EVENT RATE} 
\bigskip

 We have seen that the event rate for LSP-nucleus scattering depends on the
relative LSP-target velocity. In this section we will examine the consequences 
of the Earth's
revolution around the sun (the effect of its rotation around its axis will
be negligible) i.e. the modulation effect. This can be accomplished by
convoluting the rate with the velocity distribution.
Such a consistent choice can be a Maxwell distribution ~\cite{Jungm}
\beq
f(v^{\prime}) = (\sqrt{\pi}v_0)^{-3} e^{-(v^{\prime}/v_0)^2 }
\label{3.1}  
\eeq
provided that
\beq
v_0 = \sqrt{(2/3) \langle v^2 \rangle } =220 Km /s
\label{3.2}  
\eeq
For our purposes it is convenient to express the above distribution in the
laboratory frame, i.e.
\beq
f({\bf v}, {\bf v}_E) = (\sqrt{\pi}v_0)^{-3} 
e^{- ({\bf v}+{\bf v}_E)^2/v_0^2}
\label{3.3}  
\eeq
where ${\bf v}_E$ is the velocity of the Earth with respect to the center
of the distribution. Choosing a coordinate system in which ${\bf \hat  x}_2$ 
is the axis of the galaxy, ${\bf \hat  x}_3$ is
along the sun's direction of motion (${\bf v}_0$)
and ${\bf \hat  x}_1 = {\bf \hat  x}_2 \times {\bf \hat  x}_3$, 
we find that the position of the axis of the ecliptic is determined
by the angle $\gamma \approx 29.80$ (galactic latitude) and the
azimuthal angle $\omega = 186.3^0$ measured on the galactic plane
from the ${\bf \hat  x}_3$ axis~\cite{KVprd}.

Thus, the axis of the ecliptic lies very close to the $x_2x_3$ plane
and the velocity of the Earth is
\beq
{\bf v}_E \, = \, {\bf v}_0 \, + \, {\bf v}_1 \, 
= \, {\bf v}_0 + v_1(\, sin{\alpha} \, {\bf \hat x}_1
-cos {\alpha} \, cos{\gamma} \, {\bf \hat x}_2
+cos {\alpha} \, sin{\gamma} \, {\bf \hat x}_3\,)
\label{3.4}  
\eeq
Furthermore
\beq
{\bf v}_0 \cdot {\bf v}_1 = v_0 v_1 
\frac{cos\, \alpha}{\sqrt{1 + cot^2 \gamma \, cos^2\omega}}
\approx v_0 v_1 \, sin\, \gamma \, cos\,\alpha
\label{3.5}  
\eeq
where 
${ v}_0$ is the velocity of the sun around the center of the galaxy,
${ v}_1$ is the speed of the Earth's revolution around the sun,
$\alpha$ is the phase of the Earth orbital motion, $\alpha =2\pi 
(t-t_1)/T_E$, where $t_1$ is around second of June and
$T_E =1 year$.

The mean value of the differential event rate of Eq. (\ref {2.19}), is defined 
by

\beq
\Big<\frac{dR}{du}\Big> =\frac{\rho (0)}{m_{\chi}} 
\frac{m}{A m_N} 
\int f({\bf v}, {\bf v}_E) \mid v_z \mid \frac{d\sigma (u,\upsilon )}{du}
d^3 {\bf v} 
\label{3.6}  
\eeq
It can be more conveniently expressed as
\beq
\Big<\frac{dR}{du}\Big> =\frac{\rho (0)}{m_{\chi}} \frac{m}{Am_N} \sqrt{\langle
v^2\rangle } {\langle \frac{d\Sigma}{du}\rangle } 
\label{3.7}  
\eeq
where
\beq
\langle \frac{d\Sigma}{du}\rangle =\int \frac{ | v_z | } {\sqrt{ \langle v^2 \rangle}} 
f({\bf v}, {\bf v}_E) \frac{d\sigma (u,\upsilon )}{du} d^3 {\bf v}
\label{3.8}  
\eeq
Thus, taking the polar axis in the direction ${\bf v}_E$, we get
\beq
\langle \frac{d\Sigma}{du} \rangle = \frac{4}{\sqrt{6\pi} v_0^{4}}
\int_0^{\infty} v^3 d v \int_{-1}^{1} |\xi| d \xi 
e^{-(v^2+v_E^2+2v v_E \xi)/v_0^2 } \frac{d\sigma (u,\upsilon )}{du} 
\label{3.9}  
\eeq
or
\beq
\langle \frac{d\Sigma}{du} \rangle = \frac{2}{\sqrt{6\pi} v_E^{2}}
\int_0^{\infty} v d v \, F_0(\frac{2vv_E}{v^2_0}) \,
e^{-(v^2+v_E^2)/v_0^2 } \frac{d\sigma (u,\upsilon )}{du} 
\label{3.10}  
\eeq
with 
\beq
F_0(\chi) =\chi sinh \chi - cosh \chi + 1
\label{3.11}  
\eeq

Introducing the parameter
\beq
\delta = \frac{2 v_1 }{v_0}\, = \, 0.27,
\label{3.12}  
\eeq
expanding in powers of $ \delta $ and keeping terms up to linear in it we can
write Eq. (\ref {3.10}) as 
\barr
\langle \frac{d\Sigma}{du} \rangle & = &
                         \sigma_0 (\frac{m_x}{m_N})^2 \frac{1}{(1+\eta)^2} \,
 \{ A^2 \, [[\beta^2_0 (f^0_V - f^1_V \frac{A-2 Z}{A})^2 \bar{F}_1(u)
\nonumber \\ & + & 
(f^0_S - f^1_S \frac{A-2 Z}{A})^2 \, ]\bar{F}_0(u) -
\frac{1}{(\mu _r b)^2} \frac{2\eta +1}{(1+\eta)^2}
(f^0_V - f^1_V \frac{A-2 Z}{A})^2 u \bar{F}_0(u) ]
\nonumber \\ & + & 
(f^0_A \Omega_0(0))^2 \bar{F}_{00}(u) + 2f^0_A f^1_A \Omega_0(0) \Omega_1(0)
\bar{F}_{01}(u)  \nonumber \\ 
&+& (f^1_A \Omega_1(0))^2 \bar{F}_{11}(u) \, \} a^2
\label{3.13}
\earr
with $ \beta _0 = \upsilon _0 / c$ and
\beq
a = \frac{1}{\sqrt{2} \mu _rb\upsilon _0}  
\label{3.14}  
\eeq
The quantities 
$\bar{F}_0,\bar{F}_1,\bar{F}_{00},\bar{F}_{01},\bar{F}_{11}$  
are obtained from the corresponding form factors via the equations
\beq
\bar{F}_0(u) = F^2(u)[\Phi^{(0)}_0(a\sqrt{u})+0.135 \cos \alpha \Phi ^{(1)}_0(a\sqrt{u})]  
\label{3.15}  
\eeq
\beq
\bar{F}_{\rho ,\rho^{\prime}}(u) = F_{\rho ,\rho^{\prime}}(u)[\Phi^{(0)}_0(a\sqrt{u})+0.135 \cos \alpha \Phi ^{(1)}_0(a\sqrt{u})]  
\label{3.16}  
\eeq
\beq
\bar{F}_1(u) = F^2(u)[\Phi ^{(0)}_1(a\sqrt{u})+0.135 \cos \alpha \Phi ^{(1)}_1(a\sqrt{u})]  
\label{3.17}  
\eeq
\beq
\Phi^{(l)}_{k}(x) =  \frac{2}{\sqrt{6 \pi}} \int_x^{\infty} dy y^{2k-1}
(\exp {(-y^2)}) F_{l}(2 y ))
\label{3.18}  
\eeq
with $F_0(\chi)$ given in Eq. (\ref{3.11}) and
\beq
F_1(\chi) \, = \, 2\, \Big[ \,(\frac{\chi^2}{4} + 1) cosh\, \chi - 
\chi \,sinh \,\chi -1 \, \Big]
\label{3.19}
\eeq
 For the cases we considered in this work we find that the quantities 
$\bar{F}_{\rho ,\rho^{\prime}}(u)$ are almost the same for all isospin
channels. We believe this to be a more general result. 
The value of $0.135$ was obtained using $\sin{\gamma} \approx 0.5$

Combining Eqs. (\ref {3.7}), (\ref {3.13}) and (\ref {3.15}) - (\ref {3.18}) 
we obtain 
\beq
\Big<\frac{dR}{du}\Big> = R^0 t^0  Rr0 [1 + cos{\alpha}\, H(u)]
\label{3.20}  
\eeq
In the above expressions $R^0$ is the rate obtained in the conventional 
approach \cite {JDV} by neglecting the
momentum transfer dependence of the differential cross section, i.e. by
integrating Eq. (\ref  {3.7}) after the form factors $\bar F$ entering 
Eq. (\ref {3.13}) have been neglected. The parameter $t^0$ is the additional  
factor needed when the form factors are included and the total event rate is 
convoluted with the
velocity distribution. $Rr0$ is the relative differential rate, i.e. the
differential rate divided by the total rate, in the absence of modulation, i.e.
\beq
Rr0= \frac{1}{t^0} \frac{dr^{(0)}}{du}     
\label{3.21}  
\eeq
Note that in the above expressions $t^0$ was defined so that
the quantity $Rr0$ is 
normalized to unity when integrated from $u_{min}$ to  infinity. 
From Eqs. (\ref {3.15}) - (\ref {3.18}) we see that if we consider each
mode separately the differential modulation amplitude H takes the form
\beq
H(u) = 0.135 \frac{\Phi^{(0)}_{k}(a \sqrt{u})}{\Phi ^{(1)}_{k}(a \sqrt{u})}    
\label{3.22}  
\eeq 
Thus in this case H depends only on u and a.
This means that, if we neglect the coherent vector contribution, which,
as we have mentioned, is justified, H essentially depends 
only on the momentum transfer, the reduced mass and the size of the nucleus.

 Integrating Eq. (\ref {3.20}) we get
\beq
R = R^0 t^0  [1 +cos{\alpha}\, h(u_0,Q_{min})]
\label{3.23}  
\eeq
where $Q_{min}$ is the energy transfer cutoff imposed by the detector.
 The effect of folding
with LSP velocity on the total rate is taken into account via the quantity
$t^0$. All other SUSY parameters have been absorbed in $R^{0}$. Strictly
speaking the quantity h also depends on the SUSY parameters. It does not depend
on them, however, if one considers the scalar,spin etc. modes separately.\\

 Returning to the differential rate it is sometimes convenient, as we will see
later, to write it in a slightly different form
\beq
\Big<\frac{dR}{du}\Big> = R^0 t^0 (Rr0 + \cos{\alpha}\, Rr1)
\label{3.24}  
\eeq
Rr1 contains the effect of modulation and is given by
\beq
Rr1= \frac{1}{t^0} \frac{dr^{(1)}}{du}     
\label{3.25}  
\eeq
The meaning of Rr0 and Rr1 will become more transparent if we  
consider each mode separately.
Thus for the scalar interaction we get
 $R^0 \rightarrow R^0_{scalar}$ and
\beq
\frac{dr^{(0)}}{du} = a^2 F^2(u) \Phi ^{(0)}_0(a \sqrt{u})    
\label{3.26}  
\eeq
\beq
\frac{dr^{(1)}}{du} = 0.135\, a^2 F^2(u) \Phi ^{(1)}_0(a \sqrt{u})    
\label{3.27}  
\eeq
For the spin interaction we get a similar expression except that
$R^0\rightarrow R^0_{spin}$ and $F^2 \rightarrow F_{\rho ,\rho^{\prime}}$.
Finally for completeness we will consider the less important vector
contribution.
 We get $R^0 \rightarrow R^0_{vector}$ and
\beq
\frac{dr^{(0)}}{du} =  a^2 F^2(u) [ \Phi ^{(0)}_1(a \sqrt{u})    
                       -\frac{1}{(\mu _r b)^2} \frac{2\eta +1}{(1+\eta)^2}
                        \frac{u}{\beta _0^2}\Phi ^{(0)}_0(a \sqrt{u}) ]    
\label{3.28}  
\eeq
\beq
\frac{dr^{(1)}}{du} = 0.135  a^2 F^2(u) [ \Phi ^{(1)}_1(a \sqrt{u})    
                       -\frac{1}{(\mu _r b)^2} \frac{2\eta +1}{(1+\eta)^2}
                        \frac{u}{\beta _0^2}\Phi ^{(1)}_0(a \sqrt{u}) ]    
\label{3.29}  
\eeq 
We see that, if we consider each mode separately, $Rr0$ and $Rr1$ are
independent of all the SUSY parameters except for $m_{\chi}$. They depend
upon the nuclear physics via the relevant form factors. 

\bigskip

\bigskip
\centerline{\bf 4. RESULTS AND DISCUSSION }
\bigskip

The three basic ingredients of the LSP-nucleus scattering are 
the input SUSY parameters, a quark model for the nucleon and the structure 
of the nuclei involved. Experimentally one is interested in the differential
rate. In the present work we found it  convenient to express it in the manner
given by Eq. (\ref {3.20}), i.e. in terms of the parameters $R^0$, $t^0$, 
$Rr0$ and the convolution amplitude H.
The parameter $R^0$ contains all the information regarding the SUSY model.
It has been discussed previously (see e.g Refs. \cite {JDV,KVprd}) and it is not 
the subject of the present
work. The other parameters will be discussed below. One is also interested in 
the total rate
(see Eq. (\ref {3.23})).  For this, instead of $Rr0$ and H, one needs the 
convolution parameter h.

 The parameter $t^0$
expresses the modification of the event rate due to its dependence on the
velocity of the LSP and the folding with the LSP velocity distribution. The
obtained results, which depend on the LSP mass, the nuclear form factors and
the detector energy cutoff, Q{min}, are
presented in Tables Ia and IIa for four nuclear targets of experimental 
interest.

 The obtained results for h, the modulation of the total event rate, are shown 
in Table Ib for Pb
and in Table IIb for some other nuclei of experimental interest. We notice 
that
typically h is quite small, $\leq 5\%$. Quite surprisingly it can become much
larger for fairly light LSP and large detector energy cutoff. In other words, 
in such cases as the cutoff energy
increases the modulated amplitude decreases less than the unmodulated one.
There seems, therefore, to be a kind of trade off between the total rate
and the modulation amplitude . Thus the detector imposed cutoffs may yield a
bonus of sizable modulation effect, 
if the event rate is still detectable.

 The quantity which the experiments attempt to measure is the differential
rate. In the present work we found it convenient to work with  the relative
differential event rate with respect to the energy transfer Q, i.e.  the
differential rate divided by the total rate.  
 Instead of Q we found it convenient to express our results in terms
of the dimensionless parameter u introduced above (see Eq. (\ref {2.16})). 
 The parameter u is related to the energy transfer
by $Q=Q_{0}$ u with $Q_{0}$ given by Eq. (\ref {2.17}).   

We focused our attention on the modulation amplitude which is described either
by the parameter H (see Eq. (\ref {3.20})) or by $Rr1$ (see  eq.(\ref {3.25})). 
Rr1 and H are independent of 
the SUSY parameters and the structure of the nucleon. Rr1 
mildly depends on the nuclear structure,  
i.e. it depends
on the reduced mass of the system, the nuclear form factor and the lower energy
cutoff imposed by the detectors. H is even independent of the nuclear form
factor, but somehow it depends on the size of the nucleus.

Summarizing our results we can say the following :

\vspace {1.0 cm}
{\bf 1.  The nucleus} $_{82}Pb^{207}$ \cite{JDV,KVprd}.\\[1.0 cm]
 In this case $Q_{0}=40$ KeV. We considered both the coherent
and the spin contribution for \\

\hspace*{.5 in}$m_{\chi} =30,50,80,100,125,250,500$ GeV and $Q_{min}=0,20,40$ KeV \\

\noindent employing the harmonic oscillator form factors of Ref. \cite{KV90}. Our results
are presented in Fig. 1(a)-(j). Since the parameter H is independent of $Q_{min}$, 
it is only shown for $Q_{min}=0$. We see that H rises with u and for the same u
it decreases with the LSP mass. It can become as
large as 25 $\%$ for light LSP (see Fig. 1(c)). We notice, however, that the event
rate drops sharply after $u=0.4$, i.e. $Q=16$ KeV. Thus, the most favored region is
around $u=0.2$ or $Q=8$ KeV (see Fig. 1(b)). We also see that H is negative at
small u and becomes positive as u increases. Notice, however, that the event
rate is large at low u (see Fig. 1(a)). Hence we have cancellations in the total 
modulation amplitude. The analogous results for $Q_{min}=20$ KeV are shown in
Figs. 1(d)-(e). The latter results
are shifted compared to the previous ones by $\Delta u=0.125$ but they  appear
otherwise  similar. This is misleading since it is the  result of
the normalization adopted (the area under the curves of
Figs 1(a) and 1(d) is normalized to unity). Notice that the absolute rates are
down about a factor of 3 from those at $Q_{min}=0$. We see from Table Ia that
the total event rates are very much suppressed for $Q_{min}=40$ KeV. Thus
if such cutoffs are required by the detector, the process is unobservable.
We also present results for the spin contribution for the isospin (11)
channel in Figs. 1(f)-(g) for $Q_{min}=0$. Our results
for $Q_{min}=20$ KeV compared to those of $Q_{min}=0$ show a similar trend 
as those of Figs. 1(d-(e) when compared to those of Figs. 1(a)-(b). The other 
isospin channels show behavior similar to the (11) channel \cite{KVprd}
Thus we can say in general that the differential rate due to the spin
contribution falls quite a bit 
slower compared to the coherent rate as a function of u. We also know that the 
total rate shows a similar trend with respect to $u_{0}$ \cite {KVprd}. 
Furthermore, the quantity
$Rr1$ is a bit broader, which means that the modulation effect 
is somewhat favored in the spin contribution since a broader energy
window around the maximum can be selected.
For purposes of comparison, we present in Figs. 1(h)-(j) the analogous results for
$Q_{min}=0$ obtained for the less important coherent vector contribution.
We see that, in addition to the couplings, the LSP velocity distribution 
favors the vector contribution, but this, of course, is not enough to overcome
the suppression factor $\beta _0^2$ (see Eq. (\ref{3.13})) due to the Majorana nature of the LSP.

\vspace {1.0 cm}
{\bf 2. The nucleus} $_{53}I^{127}$.\\[1 cm] 
This nucleus is of great experimental interest \cite{Gsasso}
due to the advantages of the NaI detector. In this case $Q_{0}=60$ KeV.
We show results for the coherent scalar interaction
employing the harmonic oscillator form factors of Ref. \cite{KV90} for\\ 
\hspace*{.5 in}$m_{\chi} =30,50,80,100,125,250$ GeV and $Q_{min}=0,45$ KeV \\
Even though for $Q_{min}=45$ KeV the total rate is suppressed (see Table IIa),
for the benefit of the experimentalists we will present the corresponding   
results for the differential rate. We do not show the differential rate for
$m_{\chi}=10$ since falls off too fast as a function of u. 
 So there is no advantage in going to an energy window. Our 
results are shown in Figs. 2(a)-(c) and Figs. 2(d)-(e) for $Q_{min}=0,45$ KeV
respectively. Results for $Q_{min}=0$ KeV are also shown in Figs. 2(g)-(h)
in the case of the spin contribution for the isospin (11) channel. 
The other channels show a similar behavior.
The spin form factors were taken from Ref. \cite{Ress}.

\vspace {1.0 cm}
{\bf 3. The nucleus} $_{11}Na^{23}$.\\[1 cm] 
 This nucleus is a part of the same detector as in the previous one. Here
$Q_0=630$ KeV. Even though for this light nucleus the spin contribution may be
relatively more important compared to the coherent one, we
only considered in this work the coherent contribution in a fashion analogous to the 
Al case discussed below. The parameters $t^0$ and h are shown in
Tables IIa, IIb respectively. In this case the detector energy cutoff is
$8-16$ KeV. Our results for the differential rate for zero energy
cut off are similar to those for Al listed below. For $Q_{min}=16$ KeV they     are shown 
in Figs. 3(a)-(b).
We see that in all cases the
differential rate falls off real fast as a function of u. This is not 
surprising since for such a light system
the momentum transferred to the nucleus cannot be large.

\vspace {1.0 cm}
{\bf 4. The nucleus} $_{13}Al^{27}$.\\[1 cm] 
 A detector with this nucleus has the advantage of very low energy 
threshold $Q_{min}=0.5$ KeV. 
In this case $Q_0=480$ KeV.
 Again only the coherent scalar contribution was considered. Both
harmonic oscillator and Woods-Saxon form factors were tried. The difference
between them was small. The results presented were obtained with  the Woods-Saxon
form factors with $c=3.07$ and $a_0=0.519$ fm \cite {Lomb}.
The parameters $t^0$  and h for various LSP masses 
 and cutoffs  are given in Tables IIa and IIb respectively.
In our plots we considered the values of
 $m_{\chi} =10,20,30,50$ GeV. For larger masses the results remain unchanged.
For $Q_{min}=0.5$ KeV our results for the
differential rate  are shown in Figs. 4(a)-(c).

\bigskip
\bigskip
\centerline{\bf 4.  CONCLUSIONS }
\bigskip

 Detectable rates for the LSP-nucleus scattering for some choices 
in the allowed SUSY parameter 
space are possible \cite {KVprd}. Similar results have been obtained
in the form of scatter-plots by Arnowitt and Nath\cite {ref4} and more will 
appear elsewhere
\cite {WKV}. Since, anyway, the event rate is indeed very low, one should
try to exploit the modulation effect, i.e. the dependence of the event
rate on the motion of the Earth.

 In the present work, by convoluting the event
rate with the LSP velocity distribution we were able to obtain the annual
modulation effect, both for the coherent as well as the spin contribution. 
We were not concerned with the diurnal modulation since it is undetectable.
This was done both in the total rate as well
as in the differential rate with respect to the energy transferred to the
nucleus. 
 For the total rate we found it convenient to write our formalism in terms of
three factors (see Eq. (\ref {3.22})). The first one, $R^0$, depends on all  the relevant
SUSY parameters. It represents the total event rate, 
when the velocity dependence of the cross-section and the convolution are
neglected. The second, $t^0$, is the modification
of the event rate due to the velocity dependence of the cross-section and the
procedure of folding with the LSP velocity. The third  
is the modulation amplitude h. If one considers separately each mode 
(scalar, spin, vector coherent etc.) $t^0$ and h depend only on the LSP mass,
the nuclear form factors and $Q_{min}$. The parameter $t^0$ for various LSP
masses and a number of nuclear systems as a function of  various
detector energy cutoffs is shown in Tables Ia, IIa. 
The total modulation amplitude h is shown in Tables Ib, IIb. We see  that
it is possible to have a modulation effect which is larger than the typical
value, $h\leq 5\%$, but in those cases when the total rate is suppressed, e.g.
for relatively small LSP mass and large $Q_{min}$. So detectors with large cutoffs should not be offhand 
considered to be disadvantaged provided that the total event rate is detectable.
 
 In the case of the differential rate, in addition to the factors $t^0$ and h
mentioned above we needed two more factors (see Eq. (\ref {3.20})).  
The relative differential rate Rr0, i.e. the differential rate divided
by the total rate, and the differential modulation amplitude H. 
If one considers separately each mode, H depends only on the
reduced mass and the size of the nucleus. The differential modulated rate 
$Rr1$ depends
in addition on the nuclear form factors. It is negative at small momentum
transfer and becomes positive as the momentum transfer increases. As a result, h  
is always less than $5\% $ \cite {KVprd} and tends to decrease in the case of
heavier nuclei. This happens because ,in the case of $Q_{min}=0$, contributions from 
different regions of
the momentum transfer tend to cancel.

 Our main result is that
the differential modulation amplitude H can become quite large as the momentum
transfer increases (see Figs. 1(c), 2(c), and 4(c)). Our results are very
encouraging. Whether this nice feature,however, can be fully exploited by
the experimentalists  will depend on whether they can exploit the energy
windows around the maximum of Rr1 shown in Figs. 1(b), (e), (g), 2(b), 2(e),
3(b) and 4(b).
The vector coherent contribution, see Figs. 1(h)-(j), shows even better features
, but unfortunately it cannot be utilized, since the total rate $R^0$
associated with it is suppressed due to the Majorana nature of the LSP. 

 In any event we found many circumstances such that the modulation effect,
both in the total as well as in the differential event rate, may 
aid the experimentalists in discriminating against background.

\bigskip
\bigskip
{\it Acknowledgements:} The author would like to
acknowledge partial support of this work by $\Pi $ENE$\Delta $ 1895/95 of 
the Greek Secretariat for research, TMR Nos  ERB FMAX-CT96-0090  and  
ERBCHRXCT93-0323 of the European Union and  the Bartol Research Foundation 
where most of the  work was done. He would like also to thank Drs S. Pittel 
and Q. Shafi  for their hospitality and useful discussions. Special thanks to 
Dr T. S. Kosmas for his help in preparing the manuscript.

\newpage
\noindent {\bf Figure Captions:}

\noindent {\bf Fig. 1}: The relative differential event rate Rr0 and   
the amplitudes for modulation Rr1 and H vs u for the target 
$_{82}Pb^{207}$ (for the definitions see text). 
The curves shown
correspond  to LSP masses as follows:\\
i) Thick solid line $\Longleftrightarrow m_{\chi}=30$ GeV.
ii) Solid line $\Longleftrightarrow m_{\chi}=50$ GeV.
iii) Dotted line $\Longleftrightarrow m_{\chi}=80$ GeV.
iv) Dashed line $\Longleftrightarrow m_{\chi}=100$ GeV.
v) Intermediate dashed line $\Longleftrightarrow m_{\chi}=125$ GeV.
vi) Fine solid line $\Longleftrightarrow m_{\chi}=250$ GeV.
vii) Long dashed line $\Longleftrightarrow m_{\chi}=500$ GeV.
If some curves of the above list seem to have been ommitted, it is
understood that they fall on top of vi). Note that, due to our
normalization of Rr0, the area under the corresponding curve is unity.
 
(a)  Rr0 for the scalar contribution 
and $Q_{min}=0$ .

(b) The amplitude Rr1 for the scalar 
contribution and $Q_{min}=0$ .

(c) The modulation amplitude H, i.e. the ratio of Rr1 divided by Rr0 for $Q_{min}=0$ .

(d) The same as in (a) for $Q_{min}=20$ KeV.

(e) The same as in (b) for $Q_{min}=20$ KeV.

(f) The same as in (a) for the spin contribution
in the isospin (11) channel. For the other isospin channels the results are
similar. 

(g) The same as in (b) for the spin contribution
in the (11) channel.

(h) The same as in (a) for the vector coherent 
contribution.

(i) The same as in (b) for the vector coherent 
contribution.

(j) The same as in (c) for the vector coherent 
contribution.

\vspace {1.0 cm}
\noindent{\bf Fig. 2}: The same as in Fig. 1 for the target $_{53}I^{127}$ .

(a) Rr0  for $Q_{min}=0$.

(b) Rr1 for $Q_{min}=0$.

(c) H  for $Q_{min}=0$.

(d) The same as (a) for $Q{min}=45$ KeV.

(e) The same as (b) for $Q{min}=45$ KeV.
The style of the curves is the same as in Fig. 1.

\vspace {1.0 cm}
\noindent{\bf Fig. 3}: The same as in Fig. 1 for the target $_{13}Na^{23}$. 
The curves shown
correspond  to LSP masses as follows:\\
i) Dotted line $\Longleftrightarrow m_{\chi}=10$ GeV.
ii) Dashed line $\Longleftrightarrow m_{\chi}=20$ GeV.
iii) Long dashed line $\Longleftrightarrow m_{\chi}=30$ GeV.
iv) Fine solid line $\Longleftrightarrow m_{\chi}=50$ GeV.

 For LSP masses heavier than 50 GeV the curves
cannot be distinguished from iv).

(a) Rr0 for $Q_{min}=16$ KeV.

(b) Rr1 for $Q_{min}=16$ KeV.

\vspace {1.0 cm}
\noindent{\bf Fig. 4}: The same as in Fig. 1 for the target $_{13}Al^{27}$ 
and $Q_{min}=0.5$ KeV. (a), (b), and (c) refer to Rr0, Rr1 and H 
respectively. The style of the curves is the same as in Fig. 3.


\newpage
\begin{table}  
{\bf TABLE Ia.} The quantity $t^{0}$ for the target $_{82}Pb^{207}$.
$t^0$ takes into account the velocity
dependence of the event rate and the folding with the LSP velocity
distribution. It is computed for
various LSP masses in the allowed SUSY parameter space. The scalar, the vector
coherent (k=1) as well as the spin contributions are included. In the latter
(11),(01) and (11) indicate the possible isospin channels. 
\vskip0.2cm

\begin{center}
\begin{tabular}{|l|c|rrrrrrr|}
\hline
\hline
& & & & & & & &    \\
&  & \multicolumn{7}{|c}{LSP \hspace {.2cm} mass \hspace {.2cm} in GeV}  \\ 
\hline 
& & & & & & & &    \\
   Mode &  $Q_{min}(KeV)$  &
     30  & 50  & 80  & 100 &
   125 & 250  & 500 \\
\hline 
& & & & & & & &    \\
          &0  & 1.23 &0.728 &0.413 &0.316&0.246 &0.123 &0.0761\\ 
Scalar    &20 & 0.404 &0.331 &0.209 &0.164 &0.129 &0.0668 &0.0468\\ 
          &40 & 0  & $2\times 10^{-4}$  & $5\times 10^{-4}$ &
          $7\times 10^{-4}$ & $6\times 10^{-4}$ & $5\times 10^{-4}$ & 
          $4\times 10^{-4}$\\ 
\hline 
& & & & & & & &    \\
Vector    &0 &  3.349 &1.735 &0.902 &0.671 &0.509 &0.248 &0.151\\ 
\hline 
& & & & & & & &    \\
Spin $\qquad$ (11)  &0 &  1.57 & 1.298 &0.949 &0.793 &0.661 &0.394 &0.266\\ 
Spin $\qquad$ (11)  &20  &0.082 &0.512 &0.367 &0.344 &0.312 &0.216 &0.155\\ 
& & & & & & & &    \\
$Spin \qquad$ (00)  &0  & 1.45 & 1.13 &0.793 &0.655 &0.542 &0.318 &0.213\\ 
& & & & & & & &    \\
$Spin \qquad$ (01)  &0  & 1.51 & 1.21 &0.866 &0.719 &0.597 &0.353 &0.237\\ 
\hline 
\hline
\end{tabular}
\end{center}
\end{table}
\begin{table}  
{\bf TABLE Ib.} The same as in Table Ia for the modulation amplitude $h$.
\vskip0.2cm

\begin{center}
\begin{tabular}{|l|c|rrrrrrr|}
\hline
\hline
& & & & & & & &    \\
&  & \multicolumn{7}{|c|}{LSP \hspace {.2cm} mass \hspace {.2cm} in GeV}  \\ 
\hline 
& & & & & & & &    \\
   Mode &  $Q_{min}(KeV)$  &
     30  & 50  & 80  & 100 &
   125 & 250  & 500 \\
\hline 
& & & & & & & &    \\
          &0  & 0.0295 &0.0151 &0.0054 &0.0022&-0.0001 &-0.0005 &-0.0059\\ 
Scalar    &20 & 0.1543 &0.0774 &0.0401 & 0.0292 &0.0211 &0.0070 &0.0013\\ 
          &40 & 0.2525  &0.1598  & 0.0991 &
          0.0784 & 0.0620 & 0.0314 & 
           0.0177\\ 
\hline 
& & & & & & & &    \\
Vector    &0 & 0.0543 &0.0621 &0.0571 &0.0560 &0.0553 &0.0545 &0.0543\\ 
\hline 
& & & & & & & &    \\
Spin $\qquad$ (11) &0 &  0.0460 & 0.0307 & 0.9266 & 0.0219 & 0.0184 & 0.0113 & 0.0066\\ 
Spin $\qquad$ (11) &20  &0.1659 & 0.0926 & 0.0549 & 0.0444 & 0.0371 & 0.0234 & 0.0151\\ 
\hline 
& & & & & & & &    \\
Spin $\qquad$ (00) &0  & 0.0421 & 0.0349 &0.0238 & 0.0195 & 0.0163 & 0.0100 & 0.0056\\ 
\hline 
& & & & & & & &    \\
Spin $\qquad$ (01) &0  & 0.0440 & 0.0369 &0.0252 &0.0207 &0.0174 &0.0107 &0.0061\\ 
\hline 
\hline
\end{tabular}
\end{center}
\end{table}
\begin{table}  
{\bf TABLE IIa.} The quantity $t^0$ for the experimentally interesting
targets $ _{53}I^{127}, _{11}Na^{23}$ and $_{13}Al^{27}$ (for definitions 
see Table Ia).   
\vskip0.2cm

\begin{center}
\begin{tabular}{|l|c|rrrrrrrr|}
\hline
\hline
& & & & & & & & &   \\
&  & \multicolumn{8}{c|}{LSP \hspace {.2cm} mass \hspace {.2cm} in GeV}  \\ 
\hline 
& & & & & & & & &    \\
   Target &  $Q_{min}(KeV)$  & 
     10  & 20  & 30  & 50  & 80  & 100 &
   125 & 250  \\
\hline 
& & & & & & & & &    \\
I         &  0 &2.16 & & 1.50 & 1.04 &0.689 &0.566 &0.469 &0.287  \\
          & 20 & 0.0 & &0.089 &0.170 &0.162 &0.144 & 0.127&
           0.0855 \\
Scalar    &45&0.0& & 0.0014&0.0124 &0.0198  & 0.0201 &0.0193 &0.0150  \\
\hline 
& & & & & & & & &    \\
          &  0 & 2.13 &  & 1.40 &0.960 &0.651 &0.553 &0.473 &0.323  \\ 
Spin $\qquad$ (11) & 20 &0. 0 & &0.075 &0.153 &0.167 & 0.164 &0.158 &
            0.137  \\
                & 45 & 0.0 & &0.0018  &0.0288 &
           0.0483 & 0.0587 &0.0674 &0.0781 \\
\hline 
& & & & & & & & &    \\
Na        &  0 &2.33 & 2.32 & 2.31 & 2.30 & 2.30 & 2.30 & 2.30 & 2.30 \\
Scalar    & 8 &0.454 & 1.19 & 1.49 &1.69 & 1.69 & 1.69 & 1.69 & 1.69 \\
          & 16  &0.064 &0.570 &0.907 & 1.19 & 1.19 & 1.19 & 1.19 & 1.19 \\
\hline 
& & & & & & & & &    \\
Al        &  0 &2.32 & 2.31 & 2.30 & 2.29 & 2.29 & 2.29 & 2.29 & 2.29\\ 
Scalar    &0.5 & 2.11 & 2.22 & 2.24 & 2.25 & 2.25 & 2.25 & 2.25 & 2.25\\ 
\hline
\hline
\end{tabular}
\end{center}
\end{table}
\begin{table}  
{\bf TABLE IIb.} The same as in Table IIa for the modulation amplitude $h$.
\vskip0.2cm

\begin{center}
\begin{tabular}{|l|c|rrrrrrrr|}
\hline
\hline
& & & & & & & & &   \\
&  & \multicolumn{8}{c|}{LSP \hspace {.2cm} mass \hspace {.2cm} in GeV}  \\ 
\hline 
& & & & & & & & &    \\
   Target &  $Q_{min}(KeV)$  & 
     10  & 20  & 30  & 50  & 80  & 100 &
   125 & 250  \\
\hline 
& & & & & & & & &    \\
          &  0 &0.0508 & & .0361 & 0.0241 &0.0139 &0.0102 &0..0072 &0..0013  \\
 I        & 20 & 0.0 & &0.1298 &0.0734 &0.0426 &0.0331 & 0.0258 &
           0.0126 \\
Scalar    &45&0.0& & 0.2194 &0.1294 &0.0740  & 0.0588 & 0.0474 &0.0267  \\
\hline 
& & & & & & & & &    \\
I                  &  0 & 0.0501 &  & 0.0344 &0.0241 &0.0180 &0.0166 &0.0157 &
                          0.0149  \\ 
Spin $\qquad$ (11) & 20 &0. 0 & &0.1309 &0.0793 &0.0568 &0.0512 &0.0471 & 0.0400 \\
                   & 45 & 0.0 & &0..2215  &0.1402 &
                          0.1018 &0.0910 &0.0809 &0.0630 \\
\hline 
& & & & & & & & &    \\
Na        &  0 &0.0540 & 0.0539 & 0.0537 & 0.0535 & 0.0535 & 0.0535 & 0.0535 &
             0.0535 \\
Scalar    & 8 & 0.1334 & 0 0906 & 0.0793 & 0.0715 & 0.0715 & 0.0715 & 0.0715 &
             0.0715  \\
          &16 & 0.2039 & 0.1237 & 0.1030 & 0.0911 & 0.0911 & 0.0911 & 0.0911 &            
             0.0911 \\
\hline 
& & & & & & & & &    \\
Al        &  0 & 0.0538 & 0.0538 & 0.0596 & 0.0534 & 0.0534 & 0.0534 & 0.0534 & 
                 0.0534\\ 
Scalar    &0.5 & 0.0598 & 0.0563 & 0.0553 & 0.0545 & 0.0545 & 0.0545 & 0.0545 &        
                 0.0545\\ 
\hline
\hline
\end{tabular}
\end{center}
\end{table}
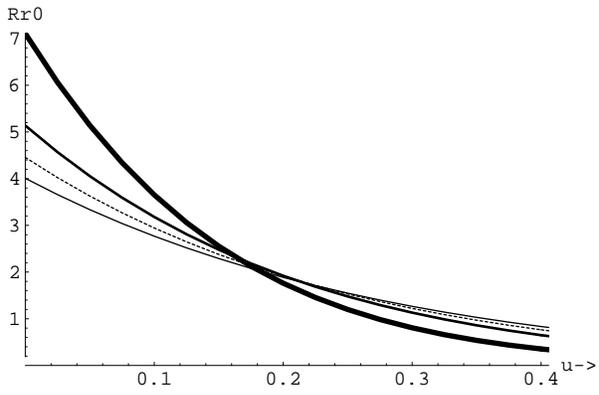
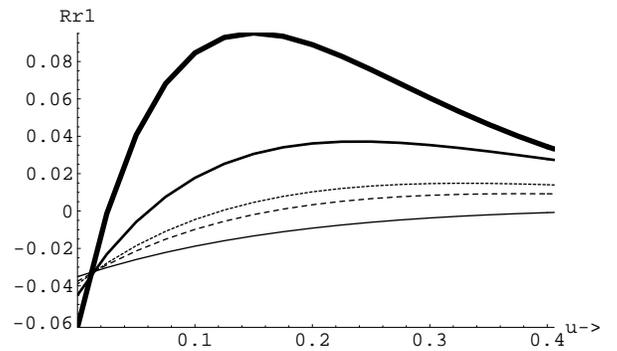
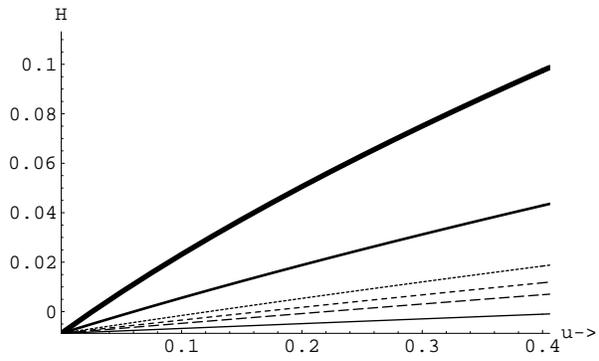
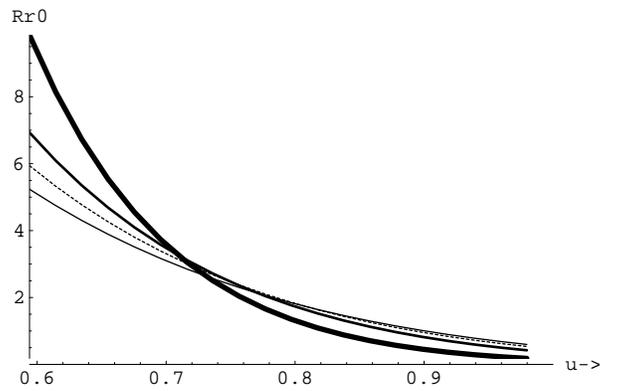
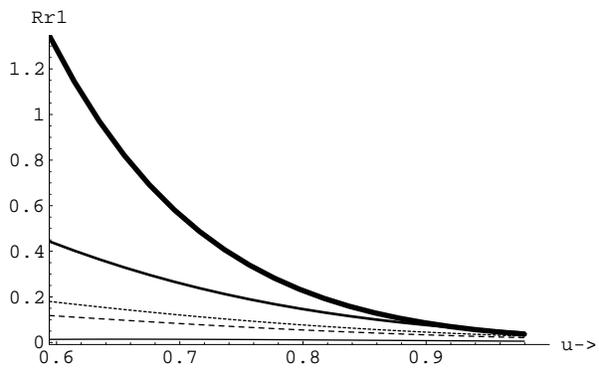
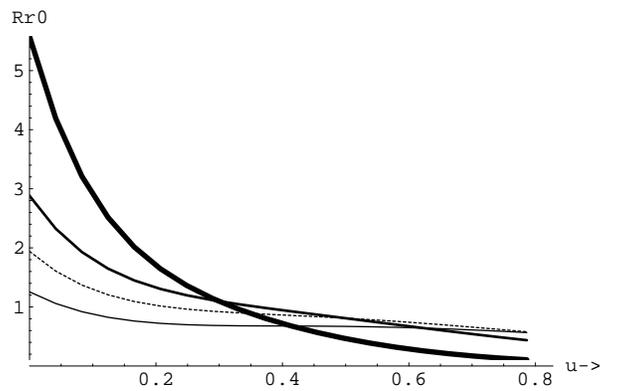
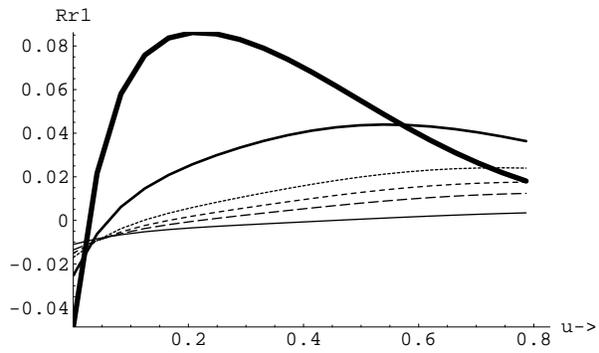
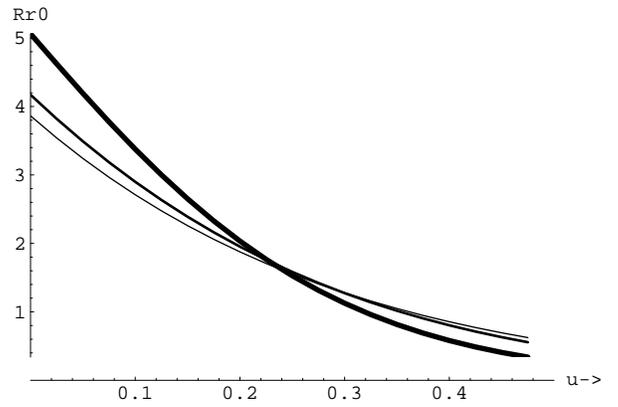
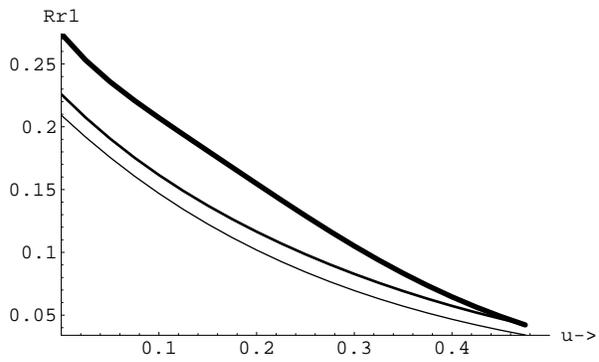
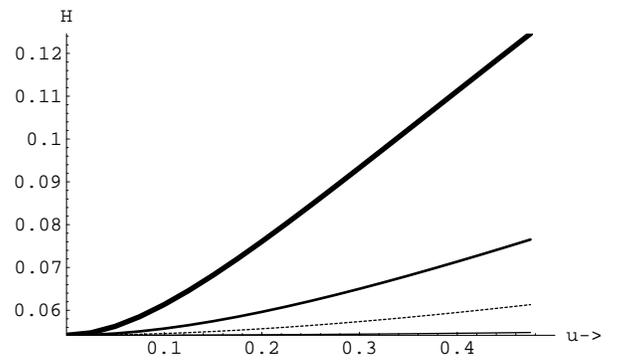
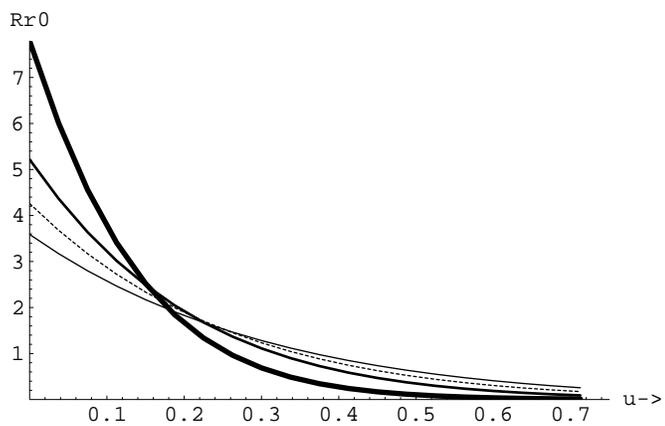
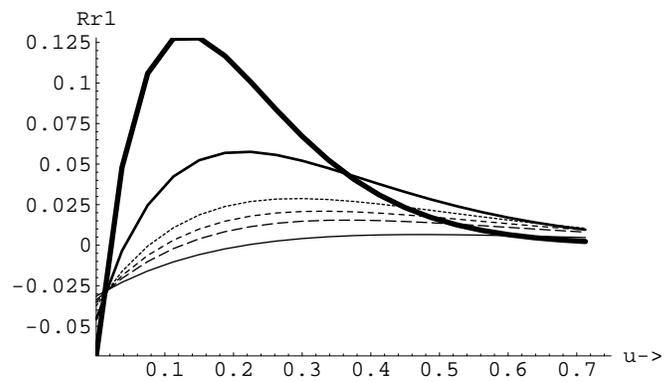
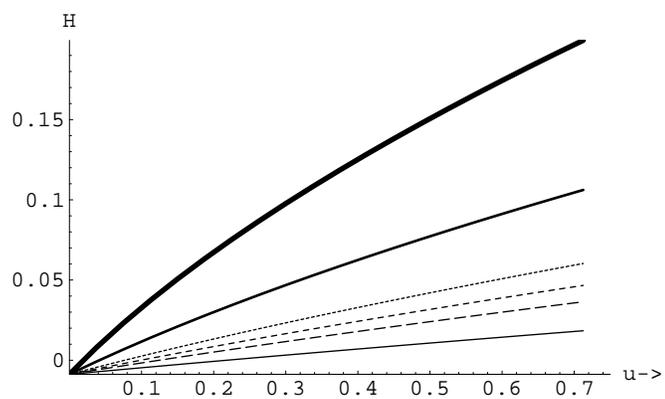
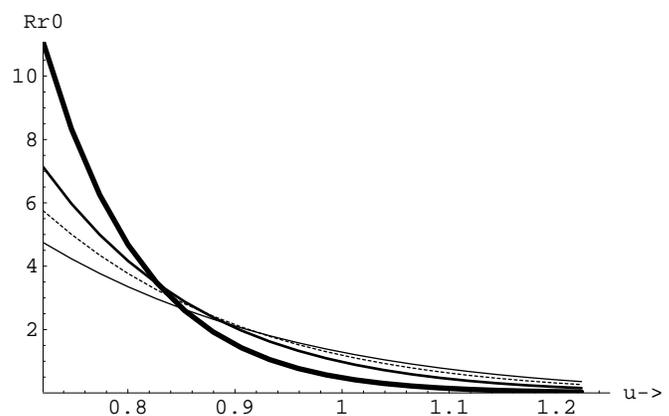
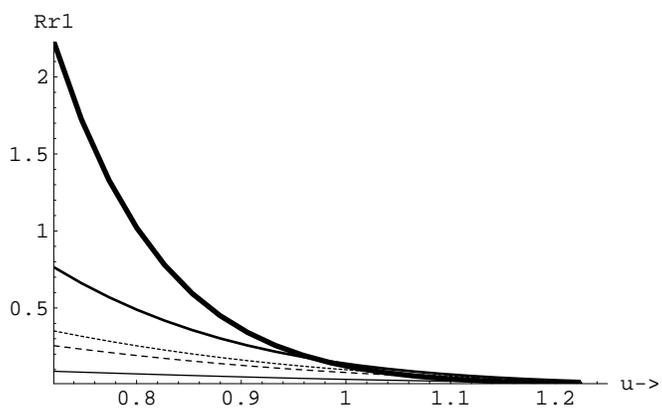
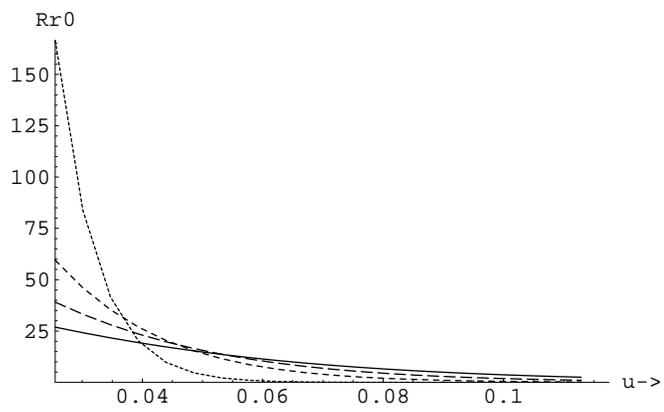
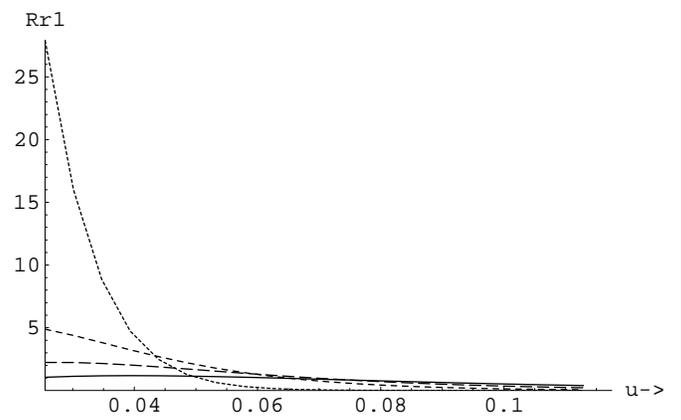
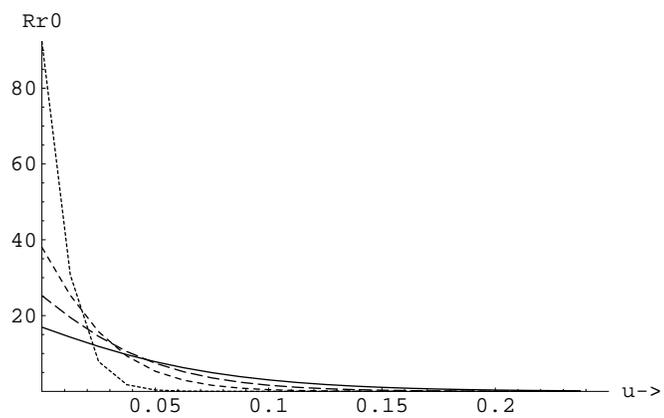
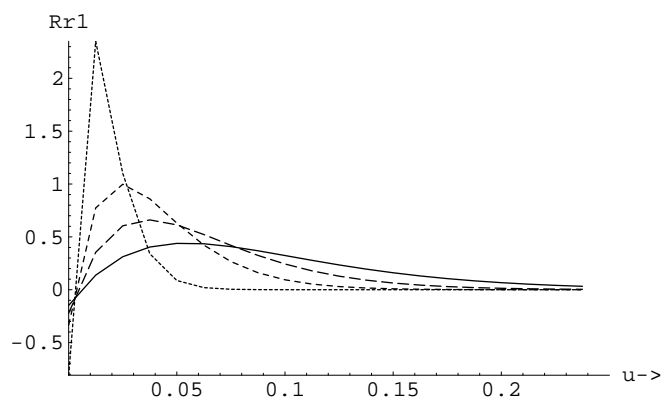
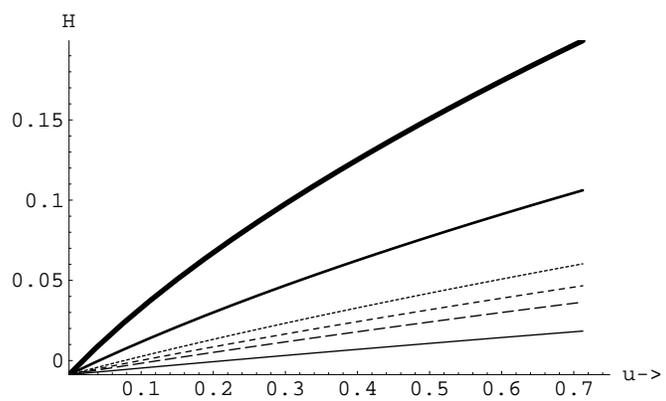
\end{document}